\documentstyle[12pt]{article}
\newcommand {\bs}{\mbox{\boldmath $\sigma $}}
\newcommand {\bB}{\mbox{\boldmath $B$}}
\newcommand {\bA}{\mbox{\boldmath $A$}}
\newcommand {\bE}{\mbox{\boldmath $E$}}
\newcommand {\bd}{\mbox{\boldmath $d$}}
\newcommand {\bk}{\mbox{\boldmath $k$}}
\newcommand {\bn}{\mbox{\boldmath $\nabla$}}
\title{ Generalized Gordon Identities, Hara Theorem and
Weak Radiative Hyperon Decays}
\author{\large E.N.~Bukina$^{1}$, 
V.M.~Dubovik$^{1}$ and V.S.~Zamiralov$^{2}$\\[3mm]
\em $^{1}$ BLTP, JINR, Dubna, Moscow reg., Russia,\\
\em e-mail: bukina@thsun1.jinr.ru, dubovik@thsun1.jinr.ru \\
\em $^{2}$ Nuclear Physics Institute, \\
\em Moscow State University,\\
\em Vorob'evy gory, 119899 Moscow, Russia, \\
\em e-mail: zamir@depni.npi.msu.su} 
\begin{document}
\maketitle
\begin{abstract}
It is shown that an alternative form of the parity-nonconserving 
(PNC) transition
electromagnetic current resolves partly a puzzle with the Hara theorem.
New formulation of it has allowed PNC weak radiative hyperon transitions 
of the charged hyperons $\Sigma^{+} \Rightarrow p + \gamma $ and
$\Xi^{-} \Rightarrow \Sigma^{-} + \gamma $
revealing hitherto unseen transition toroid dipole moment. 
\end{abstract}
%\vspace{1cm}

%%%%%%%%%%%%%%%%%%%%%%%%%%%%%%%%%%%%%%%%%%%%%%%%%%%%%%%%%%%%%%%%%%%%%%%%%%%%%%
\section{Introduction}
\setcounter{equation}{0}
%%%%%%%%%%%%%%%%%%%%%%%%%%%%%%%%%%%%%%%%%%%%%%%%%%%%%%%%%%%%%%%%%%%%%%%%%%%%%%%
$\quad$ The weak radiative decays seem to have been first analyzed theoretically
in \cite{Behr}. Unitary symmetry arrived, a theorem was proved by Hara that
decay asymmetry of the charged hyperons vanished in the exact $SU(3)_{f}$
\cite{Hara}. Since experimental discovery of a large negative asymmetry 
in the radiative decay $\Sigma^{+} \Rightarrow p + \gamma $  \cite{Ger}, 
confirmed  later \cite{PDG} (see Table 1)
, the explanation of the  net contradiction 
between  experimental results and 
the Hara theorem prediction ``has constitute a constant challenge to 
theorists'' \cite{ZenM}.  The Hara theorem was formulated at the hadron level.
But quark models while more or less succeeding in describing experimental
data on branching ratios and asymmetry parameters (see e.g. \cite{ZenM})
did not nevertheless reproduce the Hara claim without making vanish
all asymmetry parameters in the $SU(3)_{f}$ symmetry limit.
The origin of this discrepancy is not clear up to now although many
authors have investigated this problem thoroughly \cite{Ryaz}, 
\cite{Sharma}, \cite{Verma}, \cite{Zen} 
and is a real
puzzle as similar calculations say of baryon magnetic moments are known to
be rather consistent at the quark and hadron level. Also $SU(3)_{f}$
symmetry breaking effects hardly can be so large due to the well-known
Ademollo-Gatto theorem \cite{AG} to be able to account for this puzzle.
 
We shall try to show that the discrepancy between the Hara 
theorem predictions and quark model result may be overcome due to possibility
of the alternative multipole parametrization of the parity-nonconserving 
transition electromagnetic current which include not only dipole transition 
moment but also contribution of the toroid dipole moment \cite{Chesh}, 
\cite{DT}. 
Toroid dipole moment naturally
arrives in the parity-violating (PV) part of the transition radiative
matrix element and leads to reformulation of the Hara theorem. We shall
also show that Vasanti result as to the single-quark radiative transition
\cite{Vasa} is reproduced in our scheme while going from hadron to 
quark level allowing at the same time any sign of the asymmetry parameter.

%%%%%%%%%%%%%%%%%%%%%%%%%%%%%%%%%%%%%%%%%%%%%%%%%%%%%%%%%%%%%%%%%%%%%%%%%%%%%%
\section{PV electromagnetic transition current and toroid dipole moment }
\setcounter{equation}{0}
%%%%%%%%%%%%%%%%%%%%%%%%%%%%%%%%%%%%%%%%%%%%%%%%%%%%%%%%%%%%%%%%%%%%%%%%%%%%%%

$\quad$ Let us consider PV electromagnetic transition current of the two 
particles with spin and parity $1/2^{+}$.
Its possible form
is not unique as the most general expression can be written in terms of
$5$ Lorentz structures  
$ \gamma_{\mu} \gamma_{5}, \, P_{\mu}\gamma_{5},\, k_{\mu}\gamma_{5}, \,
 \sigma_{\mu \nu} k^{\nu} \gamma_{5} $ and 
$ i \epsilon_{\mu \nu \rho \lambda} \gamma_{\nu}P_{\rho}k_{\lambda} $, where 
$ P_{\mu}=(p_{1}+p_{2})_{\mu}, \, k_{\mu}=(p_{1}-p_{2})_{\mu}, \,
 \sigma_{\mu \nu}=(i/2)\left[ \gamma_{\mu},\gamma_{\nu} \right] $.
But due to electromagnetic current conservation and generalized
Gordon identities (see, e.g., \cite{Chesh})

\begin{equation}
\overline{u}_{2}\left\{i\epsilon_{\mu \nu \lambda \sigma}P_{\nu}k_{\lambda}
\gamma_{\sigma}\gamma_{5} - i\Delta m \sigma_{\mu \nu}k_{\nu}+
(k^{2}_{\lambda}\gamma_{\mu}-\hat{k}k_{\mu})\right\}\gamma_{5}u_{1}=0
\label{u2}
\end{equation}

\begin{equation}
\overline{u}_{2}\left\{-ik^{2}_{\lambda}\sigma_{\mu \nu}k_{\nu}+
\Delta m(k^{2}_{\lambda}\gamma_{\mu}-\hat{k}k_{\mu})+\left[k^{2}_{\lambda}P_{\mu}-
(k_{\nu}P_{\nu})k_{\mu}\right] \right\}\gamma_{5}u_{1}=0 
\end{equation}
where $ \Delta m= m_1-m_2 $ and $u_{1}, \, u_{2}$ are Dirac spinors of 
the baryons with masses $m_{1,2}$, this transition current
can be reduced to say one of the following forms \cite{Chesh}, \cite{DT}

\begin{equation}
J^{(A)}_{\mu}(k_{\nu})=\frac{e \eta}{(2 \pi)^{3}\sqrt{1-k^{2}_{\lambda}/M^{2}}}
\overline{u}_{2}\left[
\frac{1}{M^{2}}
(k^{2}_{\lambda}\gamma_{\mu}-kk_{\mu})G^{PV}_{1}(k^{2}_{\lambda})+
\frac{1}{M}\sigma_{\mu \nu}k_{\mu}G^{PV}_{2}(k^{2}_{\lambda})
\right]\gamma_{5}u_{1},
\label{JA}
\end{equation}

\begin{eqnarray}
J^{(A)}_{\mu}(k_{\nu})=\frac{e \eta}{(2 \pi)^{3}\sqrt{1-k^{2}_{\lambda}/M^{2}}}
\overline{u}_{2} \{
\frac{1}{M}\sigma_{\mu \nu}k_{\mu}G^{(d)}(0)+
\frac{k^{2}_{\lambda}P_{\mu}- 
(k_{\nu}P_{\nu})k_{\mu}}{M(k^{2}_{\lambda}-\Delta m^{2})}
[G^{(d)}(k^{2}_{\lambda})- \nonumber \\
-G^{(d)}(0)]+ 
 \frac{i}{M^{2}}\epsilon_{\mu \nu \lambda \sigma}P_{\nu}k_{\lambda}
\gamma_{\sigma}\gamma_{5}G^{(T)}(k^{2}_{\lambda})
\} \gamma_{5}u_{1}, \qquad
\label{JB} 
\end{eqnarray}
where $\eta = \sqrt{1- \Delta m^{2} / M^{2}}$, $M=m_{1}+m_{2}$.

(The corresponding parity-conserving (PC) current can be obtained from
Eq.(\ref{JA})
just by multiplying every structure to $\gamma_{5}$ and changing the 
superscript PV to PC.) In terms of $G_{1,2}^{PC, PV}$ the decay asymmetry
is written as \cite{Behr}

\begin{equation}
\alpha=\frac{2Re(G_{2}^{PV*}(0)G_{2}^{PC}(0))}{|G_{2}^{PV}(0)|^2
+|G_{2}^{PC}(0)|^2}.
\end{equation}

But the formfactors introduced by Eq.(\ref{JA}) do not correspond to the well-
defined multipole expansion of currents \cite{Chesh}, \cite{DT}, 
\cite{D}, \cite{Bart}, \cite{Sachs}. That is why we would like to base
our discussion on the Eq.(\ref{JB}) which as has been shown explicitly in \cite{D}
does correspond to the definite multipole expansion in a properly chosen
reference system, where $k^{2}_{\mu}=\Delta m^{2}-\bk^2$ .
This reference system is given by the equality of the
kinetic energies (e.k.e.) of the both baryons involved and enables us to
write the nonrelativistic reduction in the form \cite{D}

\begin{eqnarray}
G^{(d)}(k_{\mu}^{2})\overline{u}_{2}i\sigma_{\mu \nu}k_{\nu}\gamma_{5}u_{1}A_{\mu}
\rightarrow G^{(d)}_{e.k.e.} (\bk^2) \phi_{2}^{+} \bs \phi_{1}
[ \bE + i \Delta m \bA ]  \nonumber \\
\rightarrow \bd \bE - \dot{\bd} \bA,
\label{Gd}
\end{eqnarray}

\begin{eqnarray}
G^{(T)}(k_{\mu}^{2})\overline{u}_{2}i \epsilon_{\mu \nu \rho \lambda} \gamma_{\nu}
P_{\rho}k_{\lambda} u_{1} A_{\mu} \rightarrow G_{e.k.e.}^{(T)} (\bk^2)
\bk \times [ \bk \times \bs ] \bA \nonumber \\
\rightarrow G_{e.k.e.}^{(T)} (\bk^2) \phi_{2}^{+} \bs \phi_{1} \bn
\times \bB.
\end{eqnarray}

Here $d$ is the transition dipole moment, $\bE$ and $\bB$ are the 
electric and
magnetic fields, respectively, $\phi_{1,2}$ are Pauli spinors of the
baryons involved. 

One can see that indeed the parametrization given by the Eq.(\ref{JB}) is a
multipole one where the dipole transition and toroid dipole transition
moments are given, respectively by

\begin{eqnarray}
d=(e/M)G^{(d)}(0)=
(e/M)G_{e.k.e}^{(d)}(\Delta m^{2}) ,
\label{2.81} \\
%r^{2}_{d}=(10e/M)\frac{dG^{(d)}}{dk^{2}_{\lambda}}(0)
%\label {2.82} \\ 
T=(e/M^{2})G^{(T)}(0)=
(e/M^{2})G_{e.k.e.}^{(T)}(\Delta m^{2}) .
\label{2.83}
\end{eqnarray}

The derivatives of the formfactors $ G^{(d)}(k^{2}_{\lambda})$ and
$G^{(T)}(k^{2}_{\lambda})$ define the corresponding transition
averaged radii. As

\begin{equation}
G_{2}^{PV}(k^{2}_{\lambda})=
G^{(d)}(0)+\frac{k^{2}_{\lambda}-\Delta m^2}{M \Delta m}
G^{(T)}(k^{2}_{\lambda})
\end{equation}
we obtain that

\begin{equation}
(e/M) G_{2}^{V}(0)=d-\Delta m T .
\end{equation}

Note that $\Delta m$ here has pure kinematical origin, that is
with $\Delta m=0$ the decay discussed would not go.
This formula partly resolves a puzzle with the Hara theorem. Indeed in 
the $SU(3)_{f}$ limit:
\begin{itemize}
\item The dipole transition moments of the charged hyperon decays
should vanish and presumably stay small due to Ademollo-Gatto theorem
\cite{AG} even in the presence of the $SU(3)_{f}$ breaking terms;
\item The toroid transition dipole moments defined by the Eq.(\ref{2.83}) 
need not to be zero for these decays as their contributions decouples 
automatically in the limit $\Delta m=0 $.
\end{itemize}
So the toroid  transition dipole moment of the 
$ \Sigma^{+}\Rightarrow p+\gamma $ may be in the origin of the large asymmetry
observed \cite{PDG}.

%%%%%%%%%%%%%%%%%%%%%%%%%%%%%%%%%%%%%%%%%%%%%%%%%%%%%%%%%%%%%%%%%%%%%%%%%%%%%%
\section{The extension of the Hara theorem}
\setcounter{equation}{0}

%%%%%%%%%%%%%%%%%%%%%%%%%%%%%%%%%%%%%%%%%%%%%%%%%%%%%%%%%%%%%%%%%%%%%%%%%%%%%%c

$\quad$ In order to state our result in another way we write the PV part of the 
radiative transition matrix element  with the Lorentz structure 
$O^{T}_{\mu} = i \epsilon_{\mu \nu \lambda \rho}P_{\nu} k_{\lambda} 
\gamma_{\rho}$
in the framework of the  $SU(3)_{f}$ symmetry approach following
strictly \cite{Hara} as

\begin{eqnarray}
M=J^{(T)}_{\mu} \epsilon_{\mu} + H.C.
=\{a^{T}(\overline{B}^{2}_{3}O^{T}_{\mu}B^{1}_{1}+
\overline{B}^{3}_{2}O^{T}_{\mu}B^{1}_{1}
+\overline{B}^{1}_{1}O^{T}_{\mu}B^{2}_{3}+\overline{B}^{1}_{1}O^{T}_{\mu}B^{3}_{2})+
\nonumber \\
c^{T}(\overline{B}^{3}_{1}O^{T}_{\mu}B^{1}_{2}+\overline{B}^{2}_{1}O^{T}_{\mu}B^{1}_{3}+
\overline{B}^{1}_{2}O^{T}_{\mu}B^{3}_{1}+\overline{B}^{1}_{3}O^{T}_{\mu}B^{2}_{1})
\} \epsilon_{\mu}
\label{new}
\end{eqnarray}
where $B_{\alpha}^{\beta}$ is the $SU(3_{f})$ baryon octet, $B_{1}^{3} = p,
\, \, B_{1}^{2} = \Sigma^{+}$ etc., and $a^{T}$ and $c^{T}$ are up to 
a factor the toroid dipole moments of the neutral and charged hyperon 
radiative transitions, respectively.
                           
Here positive signs in front of every baryon bilinear combination
arrive due to Hermitian properties of the relevant Lorentz
structure.Now all $6$ PV radiative transitions are open in contrast to the
Hara result

\begin{eqnarray}
M=J^{(d)}_{\mu} \epsilon_{\mu} + H.C.
= a^{d}(\overline{B}^{2}_{3}O^{d}_{\mu}B^{1}_{1}+
\overline{B}^{3}_{2}O^{d}_{\mu}B^{1}_{1}
-\overline{B}^{1}_{1}O^{d}_{\mu}B^{2}_{3}-
\overline{B}^{1}_{1}O^{d}_{\mu}B^{3}_{2})
%\nonumber \\
\label{hara}
\end{eqnarray}
based on another Lorentz structure form 
$i \sigma_{\mu \nu} k_{\nu}\gamma_{5} $ \cite{Hara} which in turn comes to 
hadrondynamics from QED. We display in the Table 2 the results of 
Eq.(\ref{new}) and
\cite{Hara} together with the result of a traditional single-quark radiative 
transition which we have taken from \cite{Sharma}. The parameter $c^{T}\sim T$ 
in the 3rd column of the Table 2 opens
a possibility to account for large nonzero asymmetry in the charged
hyperon radiative decays even in the $SU(3)_{f}$ symmetry limit
for the corresponding coupling constants.

%%%%%%%%%%%%%%%%%%%%%%%%%%%%%%%%%%%%%%%%%%%%%%%%%%%%%%%%%%%%%%%%%%%%%%%%%%%%%%

\section{New derivation of the Vasanti formula }
\setcounter{equation}{0}

%%%%%%%%%%%%%%%%%%%%%%%%%%%%%%%%%%%%%%%%%%%%%%%%%%%%%%%%%%%%%%%%%%%%%%%%%%%%%%

$\quad$ Radiative hyperon decays were analyzed in \cite{Vasa} also 
at the quark level upon taking into account
chiral invariance considerations. We shall try to rederive 
the main result of \cite{Vasa} , namely, that the PV single-quark
radiative transition $ s \rightarrow d + \gamma $ is 
proportional to $ (m_{s}-m_{d}) $ ,
using the Lorentz structure $ O^{T} $.
At the quark level we write  for the $ s \Rightarrow d+ \gamma $
transition matrix element

\begin{equation}
M=\overline{d} \gamma_{5}(A+B \gamma_{5})i \epsilon_{\mu \nu \lambda \rho}
P_{\nu} k_{\lambda} \gamma_{\rho} \epsilon_{\mu} \gamma_{5} s
\end{equation}
and upon using Eq.(\ref{u2}), where now all quark quantities are
assumed, arrive at

\begin{equation}
M=\overline{d}\left[A(m_{s}+m_{d})+B(m_{s}-m_{d})\gamma_{5}\right]
i \sigma_{\mu \nu} k_{\nu} \epsilon_{\mu}s,
\label{M}
\end{equation}
that is in fact the main Vasanti result \cite{Vasa} is reproduced.
The factors $(m_{s}\pm m_{d})$ arrive due to the generalized Gordon
identities. The relative signs of $A$ and $B$ are not fixed here so
it is possible to obtain negative value of the asymmetry parameter. 
With the chiral invariance induced one gets exactly
the Vasanti formula \cite{Vasa} as then $A=B$. Note that Eq.(\ref{M})
(with $A=B$) was obtained in \cite{Vasa} upon assuming (i) chiral
invariance, (ii) validity of the original Hara theorem. We have proved
in fact that the introduction of the toroid structure at the quark level
is in some way 
equivalent to the chiral invariance approach of \cite{Vasa} and to
the diagram approach result of \cite{Gad}. This result dictates
the insertion of the factor $(m_{s}-m_{d})$ into the parameter $c$ 
(see the 2nd column of the Table 2, and single-quark transition terms in
\cite{Sharma}, \cite{Verma} and other works cited in \cite{ZenM}) to assure
the correct behaviour of the corresponding quark PV transition amplitudes.
And $vice \quad versa$ the results of \cite{Vasa} and \cite{Gad} together
with the generalized Gordon identities have just shown that at the 
quark level it is a toroid dipole moment which is generated with 
its characteristic Lorentz structure  
$O^{T}=i\epsilon_{\mu \nu \lambda \rho}P_{\nu}k_{\lambda} \gamma_{\rho}$ .
%%%%%%%%%%%%%%%%%%%%%%%%%%%%%%%%%%%%%%%%%%%%%%%%%%%%%%%%%%%%%%%%%%%%%%%%%%%%%%%%%%
\section{Summary and Conclusion}
\setcounter{equation}{0}

%%%%%%%%%%%%%%%%%%%%%%%%%%%%%%%%%%%%%%%%%%%%%%%%%%%%%%%%%%%%%%%%%%%%%%%%%%%%%%%%%%

$\quad$ In order to resolve a contradiction between the experiments claiming large
negative asymmetry in $ \Sigma^{+}\Rightarrow p+\gamma $, the Hara 
theorem, predicting zero asymmetry for
$ \Sigma^{+}\Rightarrow p+\gamma $ and $ \Xi^{-}\Rightarrow 
\Sigma^{-}+\gamma $
in the exact  $SU(3)_{f}$ symmetry 
and quark models which cannot reproduce the Hara theorem
results without making vanish all asymmetry parameters in the 
$SU(3)_{f}$ symmetry limit,
we have considered a parity-violating part of the transition
electromagnetic current in the alternative form allowing well-defined multipole
expansion. Part of it which is connected with the Lorentz structure
$i \epsilon_{\mu \nu \lambda \rho}P_{\nu} k_{\lambda} \gamma_{\rho} $
enables as to reformulate the Hara theorem thus opening a possibility
of nonzero asymmetry parameters for all $6$ weak radiative hyperon decays
and revealing hitherto unseen transition toroid dipole moments. 
Our result is consistent with the traditional results of the single-quark 
transition models (see column 2 of the Table 2) if the relevant parameter 
has an intrinsic kinematical factor $\Delta m$. We also have reproduced 
Vasanti formula at the quark level. 

%Our result obtained at the hadron level is consistent with  existing
%quark model results and reproduces Vasanti formula at the quark level.
%is consistent with the traditional results of the single-quark
%transition models

%\vspace{30mm}
\pagebreak
\begin{center}
{\bf Table 1.  Hyperon radiative transitions, experiment}
\vspace{10mm}

\begin{tabular}{|c|c|c|} \hline
Transition                              &  BR                &   Asymmetry           \\ \hline
$\Sigma^{+} \rightarrow p \gamma$        & $1.23 \pm 0.06$    &  $-0.76 \pm 0.08$     \\ \hline 
$\Sigma^{0} \rightarrow n \gamma$        & $-$                &       $-$             \\ \hline
$\Lambda^{0} \rightarrow n \gamma$       & $1.63 \pm 0.14$    &       $-$             \\ \hline
$\Xi^{0} \rightarrow \Lambda \gamma$     & $1.06 \pm 0.16$    &  $+0.44 \pm 0.44 $    \\ \hline
$\Xi^{0} \rightarrow \Sigma^{0} \gamma$  & $3.56 \pm 0.43$    &  $+0.20 \pm 0.32 $    \\ \hline
$\Xi^{-} \rightarrow \Sigma^{-} \gamma$  & $0.128 \pm 0.023$  &  $+1.0 \pm 1.3$       \\ \hline
\end{tabular}
\end{center}

\vspace{5mm}

\begin{center}
{\bf Table 2.  Hyperon radiative transitions, theory}
\vspace{10mm}

\begin{tabular}{|c|c|c|c|} \hline
Transition PNC amplitude  &  in \cite{Sharma} & in \cite{Hara} 
& from Eqs.(\ref{new})  \\ \hline
$ \Sigma^{+}\rightarrow p \gamma$  & $-\frac{1}{3}b$ & $0$ & $c^{T}$  \\ \hline 
$ \Sigma^{0}\rightarrow n \gamma$  & $\frac{1}{3\sqrt{2}}b  $
& $\frac{1}{\sqrt{2}}a^{d} $
& $\frac{1}{\sqrt{2}}a^{T} $  \\ \hline
$\Lambda^{0}\rightarrow n \gamma$  & $\frac{3}{\sqrt{6}}b$ 
& $\frac{1}{\sqrt{6}} a^{d} $
& $\frac{1}{\sqrt{6}}a^{T} $ \\ \hline
$\Xi^{0}\rightarrow \Lambda \gamma$  & $\frac{1}{\sqrt{6}}b$ 
& $- \frac{1}{\sqrt{6}} a^{d} $
& $\frac{1}{\sqrt{6}}a^{T} $ \\ \hline
$\Xi^{0}\rightarrow \Sigma^{0} \gamma$  & $ -\frac{5}{3\sqrt{2}}b$ 
& $ -\frac{1}{\sqrt{2}}a^{d} $
& $ \frac{1}{\sqrt{2}}a^{T}$ \\ \hline
$ \Xi^{-}\rightarrow \Sigma^{-} \gamma$  & $\frac{5}{3}b$ & $0$ 
& $ c^{T}$ \\ \hline
\end{tabular}
\end{center}

\end{document}